# Heat transport in superionic materials via machine-learned molecular dynamics

Wenjiang Zhou[1], Benrui Tang[2], Zheyong Fan[2], Federico Grasselli[3, 4],

Stefano Baroni[5, 6*], and Bai Song[1, 7*]

[1]*School of Mechanics and Engineering Science, Peking University, Beijing 100871, China.*

[2]*College of Physical Science and Technology, Bohai University, Jinzhou 121013, China.*

[3]*Department of Physics, Informatics and Mathematics, Università degli Studi di Modena e Reggio Emilia, Modena 41125, Italy.*

[4]*CNRNanoS3, Modena 41125, Italy.*

[5]*SISSA—Scuola Internazionale Superiore di Studi Avanzati, Trieste 34136, Italy.*

[6]*CNR-IOM—Istituto Officina Materiali, DEMOCRITOS SISSA Unit, Trieste 34136, Italy.*

[7]*National Key Laboratory of Advanced Micro and Nano Manufacture Technology, Peking University, Beijing 100871, China.*

[*]Corresponding author. Email: baroni@sissa.it (S. Baroni), songbai@pku.edu.cn (B. Song);

**Abstract**:

Precise modeling and understanding of heat transport in the superionic phase are of great interest. Although simulations combining Green-Kubo (GK) molecular dynamics with machine-learned potentials (MLPs) stand as a promising approach, substantial challenges remain due to the crucial impact of atomic diffusion. Here, we first show that the thermal conductivity ($\kappa$) of superionic materials calculated via conventional GK integral of the energy flux varies notably with the MLP model. Subsequently, we highlight that reliable, model-independent $\kappa$ values can be obtained by applying Onsager's reciprocal relations to correctly capture the coupled heat and mass transport. Remarkably, an anomalously invariant $\kappa$ can be observed over a wide temperature range, distinct from the characteristic trends in traditional crystals and glasses. In addition, we illustrate that conventional $\kappa$ decompositions into kinetic, potential, and cross terms suffer from ambiguities in the physical interpretation, despite their mathematical rigor. Finally, we propose a criterion for the necessity of the Onsager correction and reveal the underlying mechanism as a competition between thermally and chemically driven ion fluxes.

The superionic phase is a unique intermediate state between crystalline solids and liquids [1-3], with a subset of ions exhibiting liquid-like mobility and diffusing almost freely through a rigid lattice formed by other ions that are localized and only vibrate around their equilibrium positions [3]. This intriguing phase has been observed in a variety of materials such as ice [4, 5], iron alloys [6], and helium-water compounds [7], and is attracting rapidly growing interest as a playground for fundamental research. In addition, superionic materials also hold great application potentials. Their high ionic conductivity renders them promising candidates for next-generation energy storage technologies such as solid-state battery [8-10]. Further, their generally low thermal conductivity [11, 12], in conjunction with a relatively large electrical conductivity, makes them suitable for use in high-efficiency thermoelectric devices [12-14].

Key to the application of superionic materials is their thermal conductivity ($\kappa$). For thermoelectrics, $\kappa$ largely determines the energy conversion efficiency [14-16]. For solid-state batteries, both the thermal management and modeling depend crucially on an accurate knowledge of $\kappa$ of the electrolytes [17, 18]. Furthermore, $\kappa$ of the superionic phase is inherently coupled with ion conduction [19, 20]. However, the demand for very high pressures and temperatures presents a formidable obstacle to measuring some of the materials [6, 7], while those that could be measured often yielded anomalously temperature dependences [21, 22]. Therefore, it is of fundamental importance to precisely model and understand thermal transport in the superionic phase. Unfortunately, the coexistence of solid- and liquid-like subsystems complicates first-principles calculations of $\kappa$ [23-25]. Alternatively, Green-Kubo (GK) molecular dynamics (MD) [26, 27] empowered by machine-learned potentials (MLPs) [28-31] has emerged as a promising approach, which delivers quantum-mechanical accuracy at an affordable computational cost.

Within the conventional framework, $\kappa$ is proportional to the GK integral of the energy flux [26, 27] defined as [32]:

$$\mathbf{J} = \mathbf{J}_{\text{kin}} + \mathbf{J}_{\text{pot}} = \frac{1}{V}\left(\sum_i \mathbf{v}_i \epsilon_i + \sum_i \mathbf{w}_i \cdot \mathbf{v}_i\right). \quad (1)$$

Here, $\mathbf{J}_{\text{kin}}$ and $\mathbf{J}_{\text{pot}}$ represent the kinetic and potential terms, respectively, which are also referred to as the convective and virial terms; and $\epsilon_i = m_i \mathbf{v}_i^2/2 + U_i$ denotes the atomic energy where $\mathbf{v}_i$, $m_i$, $U_i$, and $\mathbf{w}_i$ are the velocity, mass, site potential, and per-atom virial of atom $i$, respectively. The kinetic term $\mathbf{J}_{\text{kin}}$ is particularly important for heat conduction in superionic materials [20, 33]. However, a significant challenge arises: The site potential $U_i$ is not uniquely defined in different MLP models even though the total energy remains consistent [34-36]. Furthermore, there is an additional non-uniqueness in the density-functional theory-based datasets used to train the MLPs [37], which occurs when the exchange-correlation functional is altered. These inherent ambiguities will inevitably limit the accuracy of the calculated $\kappa$ values.

Here, we explore thermal transport in two representative superionic materials, $\alpha$-$Li_3PS_4$ and $\beta$-$Cu_{1.98}Se$, by combining GK molecular dynamics with MLPs. We first show that the $\kappa$ computed via the conventional GK integral is strongly affected by the specific definition of the atomic site energy and thus by the choice of the MLP models. This issue is attributed to the non-unique projection of the total energy onto individual atoms. Subsequently, by incorporating the Onsager reciprocal relations to properly account for heat-mass coupling, we obtain reliable and model-independent $\kappa$ values. Remarkably, a nearly invariant $\kappa$ of $\alpha$-$Li_3PS_4$ is observed over a wide temperature range, distinct from traditional crystals and glasses. Finally, we discuss the decomposition of $\kappa$ into kinetic, potential, and cross terms, and explore when the Onsager correction becomes necessary.

To begin with, we present a theoretical framework for calculating the thermal conductivity of superionic materials, which builds upon the conventional Green-Kubo theory [26, 27] but incorporates Onsager reciprocal relations [38, 39] to describe the coupled heat and mass transport [19, 20]. For a system with $M$ distinct components (ions) in motion, there are $M$ corresponding conserved fluxes $J_i$: One energy flux (denoted by subscript 0) and $M-1$ mass fluxes. Any conserved flux can be expressed as a linear combination of the thermodynamic affinity $F$ and the Onsager coefficient matrix $\mathbf{\Lambda}$ [38, 39]: $J_i = \sum_{j=0}^{M-1} \Lambda_{ij} F_j$. Each component $\Lambda_{ij}$ is obtained via the GK integral as [26, 27]:

$$\Lambda_{ij} = \frac{V}{k_{\mathrm{B}}} \int_0^{\infty} \langle J_i(\mathrm{t}) \cdot J_j(0) \rangle \mathrm{d}t, \tag{2}$$

where $J_i(\mathrm{t}) \cdot J_j(0)$ is the time correlation of the fluxes and $\langle \cdot \rangle$ denotes ensemble average. In a single-component system, there is no mass flux and no contribution from the kinetic term $\mathbf{J}_{\mathrm{kin}}$ [40, 41]. Therefore, the thermal conductivity is simply given by the conventional GK formula as $\kappa = \Lambda_{00}/T^2$, where $T$ is the absolute temperature. This expression holds regardless of how the atomic energy (especially the site potential) is defined, which is known as the gauge invariance of thermal transport [42, 43].

In superionic materials with at least one moving component, $\kappa$ is defined as the ratio of the energy flux to the temperature gradient under the condition that all other conserved fluxes are zero [20], and can be derived as $\kappa = 1/[T^2(\mathbf{\Lambda}^{-1})_{00}]$. Specifically, in a two-component system, this simplifies to $\kappa = (\Lambda_{00} - \Lambda_{01}^2/\Lambda_{11})/T^2$. By introducing the notation $L_{ij} = \Lambda_{ij}/T^2$, $\kappa$ can be concisely rewritten as:

$$\kappa = L_{00} - L_{01}^2/L_{11}. \tag{3}$$

Here, $L_{00}$ is obtained through a direct GK integral of the energy flux which yields the $\kappa$ of a single-component system, $L_{01}$ describes the heat and mass (ion) coupling, and $L_{11}$ is essentially equivalent to the self-diffusion coefficient of the superionic ions and therefore the ionic conductivity. The subtracted term $L_{01}^2/L_{11}$ represents the Onsager correction arising from mass diffusion in a multi-component system. As we show later, Eq. (3) satisfies both the gauge invariance principle [42] and the convective invariance principle [20], while $L_{00}$ alone does not. Furthermore, based on the individual components of the energy flux in Eq. (1), the corresponding $L_{00}$ can be decomposed into kinetic, potential, and kinetic-potential cross terms as: $L_{00} = \kappa_{\mathrm{k}} + \kappa_{\mathrm{p}} + 2\kappa_{\mathrm{kp}}$. This decomposition has been widely employed to analyze thermal transport in superionic materials [11, 21, 36, 44-47] and will be revisited below.

Using the formalism described above, we first explore heat transport in $\alpha$-$Li_3PS_4$, a promising solid-state electrolyte with a high ionic conductivity [48]. The training datasets are obtained from a previous study [49] which performed density-functional theory (DFT) calculations for various phases of $Li_3PS_4$ at the PBEsol level [50]. Specifically, we generate six pairs of datasets by randomly partitioning the original data into separate training and test datasets. Correspondingly, six MLP models are trained using the neuroevolution potential (NEP) framework [28, 51, 52], which are denoted as models A to F. All training details are provided in the Supplemental Material [53].

With the trained MLPs, we perform MD simulations using the GPUMD package [52, 54]. In Fig. 1(a), we plot the calculated mean square displacement (MSD) of different species in $\alpha$-$Li_3PS_4$ from 350 to 850 K. At relatively low temperatures, $\alpha$-$Li_3PS_4$ behaves like an ordinary solid, as indicated by the finite MSDs of all the atoms. However, as $T$ increases beyond 350 K, the MSD of Li ions begins to diverge with the simulation time, while the MSDs of P and S atoms remain finite, even at 650 K. This behavior indicates the transition of $\alpha$-$Li_3PS_4$ into a two-component superionic phase, with $Li^+$ diffusing within the solid framework formed by $PS_4^{3-}$. This is further confirmed by the MD trajectories visualized in Fig. 1(a).

We further examine the projection of the total energy onto individual atoms, using models A and B as illustrative examples. As shown in Fig. 1(b), for the same $\alpha$-$Li_3PS_4$ structure, while the difference in total energy is remarkably small (<0.3%) between the two MLP models, the distribution and magnitude of site potentials for the Li and S atoms differ significantly. This observation also holds for other models as evidenced by the variations of the average site potential for Li (Table S3). The inherent non-uniqueness in decomposing the total energy strongly affects the simulation of thermal transport in superionic materials, which we demonstrate later.

To calculate the thermal conductivity $\kappa = L_{00} - L_{01}^2/L_{11}$ of $\alpha$-$Li_3PS_4$, we first conduct equilibrium MD simulations using the six MLP models to obtain the energy and mass fluxes. Then, the $L_{00}$, $L_{01}$, and $L_{11}$ are computed with our in-house GK code [53]. Additionally, we also calculate $\kappa = L_{00} - L_{01}^2/L_{11}$ by performing multivariate cepstral analysis (MCA) using the SporTran code [20, 55] to provide a benchmark. In Fig. 2(a), we plot representative $\kappa$ values obtained at 650 K via different approaches. First, our GK code is validated by yielding essentially the same $L_{00} - L_{01}^2/L_{11}$ as that from MCA, using any one of the six MLP models. Moreover, the $L_{00} - L_{01}^2/L_{11}$ values are highly consistent across different models and agree well with previous reports [34, 49], serving as a numerical manifestation of the convective invariance principle [20]. In the Supplemental Material, we further provide a mathematical proof [53].

Compared to $\kappa = L_{00} - L_{01}^2/L_{11}$ which accounts for the coupled heat and mass flow, the widely used expression of $\kappa = L_{00}$ from conventional GK integral of the energy flux yields values that vary significantly across the six MLP models. With models A, E, and F, the calculated $L_{00}$ values closely match $L_{00} - L_{01}^2/L_{11}$. However, with models B, C, and D, $L_{00}$ is notably larger, reaching up to a factor of three. This model dependence renders the use of $\kappa = L_{00}$ problematic for superionic materials and highlights the necessity of incorporating the Onsager correction. Interestingly, the fact that half of the models yield fairly good predictions with $\kappa = L_{00}$ suggests that this problem can easily go unnoticed. To gain insight into the underlying mechanism, we further plot $L_{01}$ and $L_{11}$ in Fig. 2(b), observing that the former follows a trend similar to that of $L_{00}$ while the latter remains consistent across different models. This indicates that the definition of atomic energy is key since it affects both $L_{00}$ and $L_{01}$, while $L_{11}$ only depends on the mass flux. Note that although different exchange-correlation functionals can lead to distinct DFT-calculated total and projected energies, only $\kappa = L_{00}$ may be affected while $\kappa = L_{00} - L_{01}^2/L_{11}$ remains robust.

The temperature dependence of $\kappa$ provides valuable insights into the heat conduction mechanism [56-58]. In Fig. 3(a), we plot both the calculated $L_{00}$ and $L_{00} - L_{01}^2/L_{11}$ as a function of temperature, using models A, C, and D as examples. At low temperatures where superionic diffusion is suppressed [Fig. 1(a)], the contribution of $\mathbf{J}_{kin}$ and the effect of heat-mass coupling are small [40, 41]. Consequently, the $L_{00}$ values from all three models are similar, thus explaining its wide and often successful adoption as the $\kappa$ of various solids. As $T$ rises, however, the three models exhibit distinct thermal transport behaviors: $L_{00}$ increases rapidly for models C and D, while rising only slightly for model A. In contrast, $L_{00} - L_{01}^2/L_{11}$ yields very consistent values and a temperature-invariant trend across all models from 300 to 850 K. This behavior is highly unusual, differing both from the classical $T^{-1}$ dependence of $\kappa$ in typical crystals [59-62] and the gradual increase of $\kappa$ with $T$ in amorphous materials [58, 63, 64], as visualized in Fig. 3(b). Interestingly, a recent study also reported a temperature-invariant $\kappa$ in meteoritic silica although only from 80 to 380 K [65]. Additionally, we verify that this $\kappa$ trend is not an artifact due to the missing nuclear quantum effects in classical MD (Fig. S14).

To understand the unusual transport behavior, we further decompose $L_{00}$ into $\kappa_k$, $\kappa_p$, and $\kappa_{kp}$ as mentioned earlier [Fig. 3(c)]. This approach has been widely employed to analyze why the $\kappa$ of superionic conductors deviate from the $T^{-1}$ trend. At low temperatures, the potential term $\kappa_p$ dominates and remains consistent across all three MLP models, while the kinetic term $\kappa_k$ and the cross term $\kappa_{kp}$ are nearly zero. However, as $T$ rises to the superionic range, a striking phenomenon is observed: $\kappa_k$, $\kappa_p$, and $\kappa_{kp}$ all exhibit pronounced model-dependent variations. These results indicate that, although the energy flux-based decomposition is mathematically valid, it is not gauge invariant and lacks a clear physical interpretation. Therefore, a precise mechanistic understanding based on such decomposition may remain elusive, despite its qualitative appeal. Alternatively, an intuitive picture may be found in the collective effect of propagating phonons and tunneling diffusion transport [65-67]. With rising temperature, the contribution of the former decreases whereas that of the latter increases, which yields a total $\kappa$ that is temperature insensitive (Fig. S16).

In addition to $\alpha$-$Li_3PS_4$, we consider $\beta$-$Cu_{1.98}Se$ which exemplifies the superior thermoelectric potential of superionic materials [12-14]. At temperatures above 400 K [36], the Cu ions in this material exhibit liquid-like diffusion, while the Se ions vibrate around their equilibrium positions. We first construct four training datasets by employing VASP [68, 69], obtaining four MLP models for subsequent MD simulations of thermal transport from 600 to 1200 K [53]. As an example, Fig. 4(a) presents the computed $L_{00}$ and $L_{00} - L_{01}^2/L_{11}$ values at 1000 K. As expected, the latter shows good consistency across all four MLP models and agrees well with the experimentally measured $\kappa$ minus the estimated electronic contribution [14]. Interestingly, in contrast to $\alpha$-$Li_3PS_4$, $L_{00}$ only shows a small model-induced variation of 8% and remains close to the experimental result. We further look into $L_{01}$ and $L_{11}$. As shown in Fig. 4(b), $L_{01}$ also exhibits similar model dependence as $L_{00}$, whereas $L_{11}$ shows a different trend and smaller variations. Despite the negligible model dependence of $L_{00}$, its decomposition into $\kappa_k$, $\kappa_p$, and $\kappa_{kp}$ still appears rather arbitrary [Fig. 4(c)].

Compared to $\beta$-$Cu_{1.98}Se$, the substantially stronger model dependence of $\kappa = L_{00}$ for $\alpha$-$Li_3PS_4$ motivates us to explore the underlying physical mechanism. Intuitively, since there is no model dependence for $\kappa = L_{00} - L_{01}^2/L_{11}$ regardless of the material, the key must lie in the Onsager correction term $L_{01}^2/L_{11}$. The calculated $L_{01}$ for $\beta$-$Cu_{1.98}Se$ is on the same order of magnitude as that for $\alpha$-$Li_3PS_4$, however, the value of $L_{11}$ is larger by a factor of 30. This leads to a much smaller correction for $\beta$-$Cu_{1.98}Se$ and therefore weaker model dependence.

We then proceed to identify simplified conditions under which the Onsager correction becomes necessary. First, to capture the heat-mass coupling, we employ the site potential $U$ of the superionic species instead of $L_{01}$, since the former can be readily obtained from the trained MLP model while the latter requires expensive additional MD simulations. As jointly demonstrated by Fig. 2 and Table S3, a smaller magnitude of $U$ always corresponds to a smaller $L_{01}$, both indicating weaker coupling. Further, $L_{11}$ can be obtained either from existing experimental data or from cost-effective simulations such as via the slope of MSD as shown in Fig. 1(a). These considerations lead to the introduction of a characteristic parameter $U^2/L_{11}$. As shown in Fig. 5, we find that the relative error $\delta$ of the thermal conductivity calculated via the conventional GK approach exhibits a log-linear relationship with $U^2/L_{11}$. Empirically, when $U^2/L_{11} < 10^{14}$ $eV^2kg^{-1}s^{-1}m^3K$, the error of using $\kappa = L_{00}$ is less than 10%.

Finally, we offer a physical explanation for why the Onsager correction diminishes with increasing $L_{11}$. To measure $\kappa$, a temperature gradient must be applied, which not only induces a heat flux (Fourier's law) but also a mass flux (Soret effect). The thermally-driven mass flux then generates a concentration or chemical potential gradient. This in turn drives a mass flux in the opposite direction (Fick's law) and also a heat flux (Dufour effect) which partially cancels the primary energy flux. This process continues until the net mass flux becomes zero. With this picture, it is straightforward to see that a larger $L_{11}$ implies a smaller concentration gradient at steady state and therefore a smaller correction to the energy flux and $\kappa$.

In summary, we have investigated thermal transport in superionic materials, demonstrating the inadequacy of the conventional Green-Kubo approach and the necessity of the Onsager correction for precise modeling. Our findings highlight the superionic phase as an intriguing playground for exploring new thermal physics.

## ACKNOWLEDGMENTS

We acknowledge Zezhu Zeng for helpful discussions on the Green-Kubo calculations. W.Z and B.S was supported by the Science Fund for Creative Research Groups from the National Natural Science Foundation of China (Grant No. 52521007), the Scientific Research Innovation Capability Support Project for Young Faculty (ZYGXQNJSKYCXNLZCXM-E1) from the Ministry of Education of China, the National Key R&D Program of China (Grant No. 2024YFA1207900), and the High-performance Computing Platform of Peking University. W.Z. acknowledges support from the China Association for Science and Technology. B.T. and Z.F. were supported by the Science Foundation from Education Department of Liaoning Province (No. LJ232510167001). B.S. acknowledges support from the New Cornerstone Science Foundation through the XPLORER PRIZE.

## DATA AVAILABILITY

The training dataset of $\alpha$-$Li_3PS_4$ is available at [70]. The training dataset, MLP models, as well as the main input and output of the MD simulations for $\beta$-$Cu_{1.98}Se$ will be made freely available at [71]. The source code for Green-Kubo calculations also will be made freely available at [71]. Other data that support the findings of this work are available from the authors upon reasonable request.

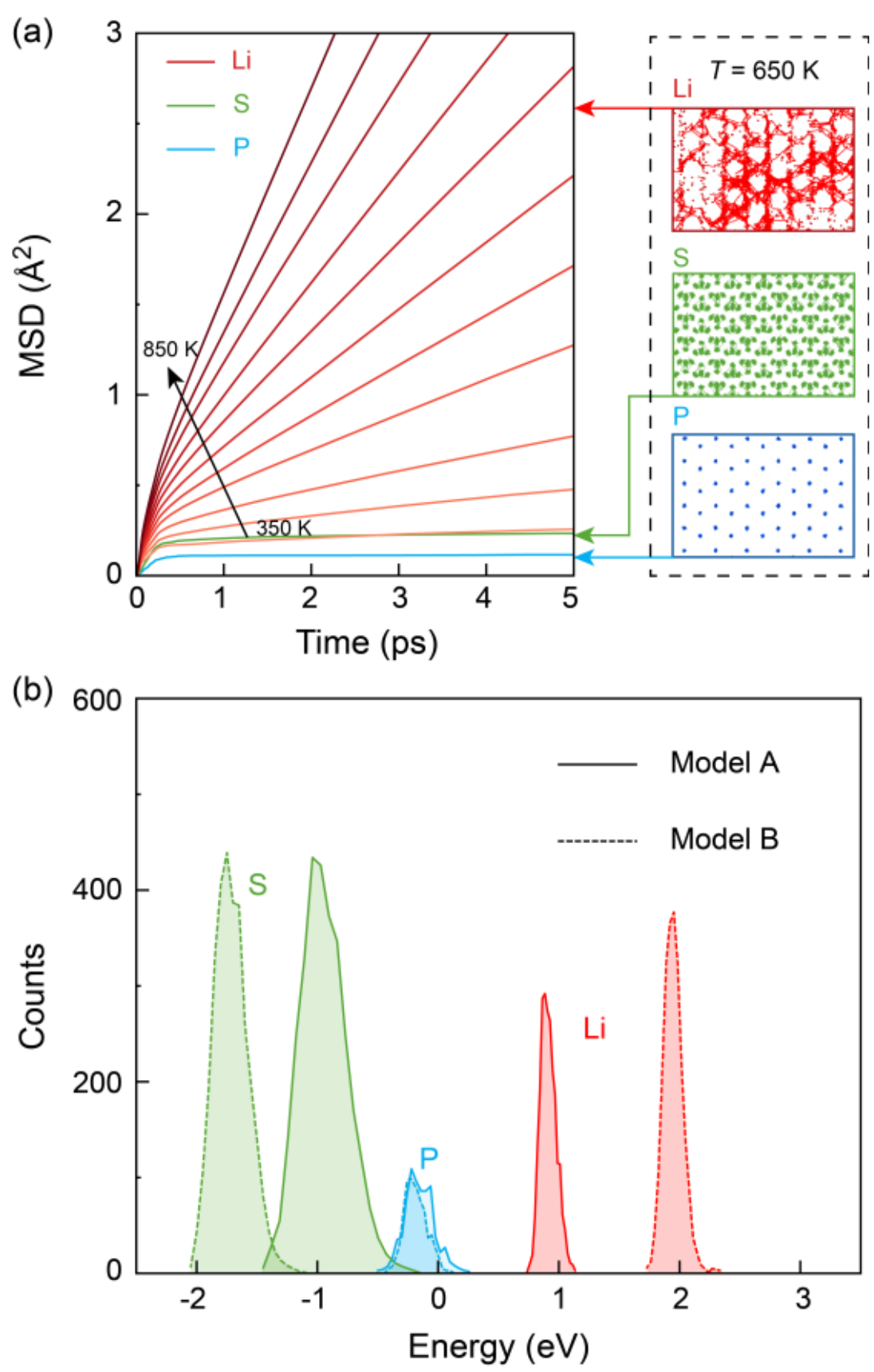

**FIG. 1. Mean square displacement (MSD) and non-uniqueness of atomic projection of total energy in $\alpha$-$Li_3PS_4$.** (a) Calculated MSD of Li, P, and S atoms in $\alpha$-$Li_3PS_4$ as a function of time and temperature. Results from 350 K to 850 K are shown for Li (red) in steps of 50 K; while for S (green) and P (blue), only the 650 K data are plotted for clarity. Right panel visualizes MD trajectories of the Li, S, and P atoms at 650 K from 1000 MD frames. Dots mark the atoms while lines trace their movement. For Li, the trajectories of only four atoms near the corner and one near the center are displayed. (b) Distribution of atomic energies in $\alpha$-$Li_3PS_4$ for the same supercell structure. Results computed from two machine-learned models (A and B) are shown as an example.

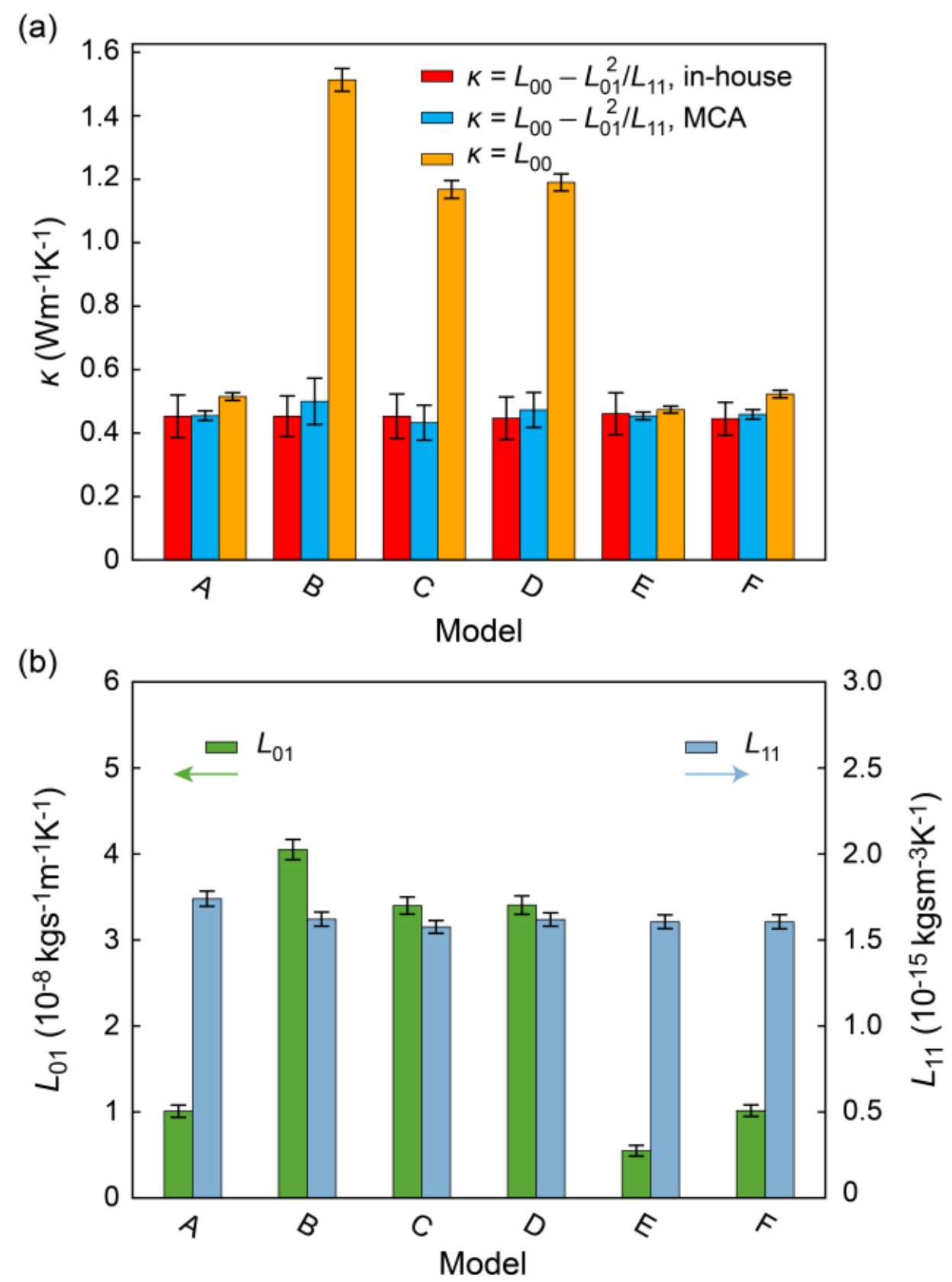

**FIG. 2. Thermal conductivity and Onsager coefficients of $\alpha$-$Li_3PS_4$ at 650 K.** (a) Thermal conductivity values obtained from the conventional Green-Kubo integral based on $L_{00}$ (orange) and also the formula $L_{00} - L_{01}^2/L_{11}$ which satisfies the invariance principle (red and blue). Here, the results from multivariate cepstral analysis (MCA, blue) are shown for comparison and validation [20]. Six different machine-learned models are considered, which are denoted as models A to F. (b) $L_{01}$ (green, left axis) and $L_{11}$ (blue, right axis) from different models.

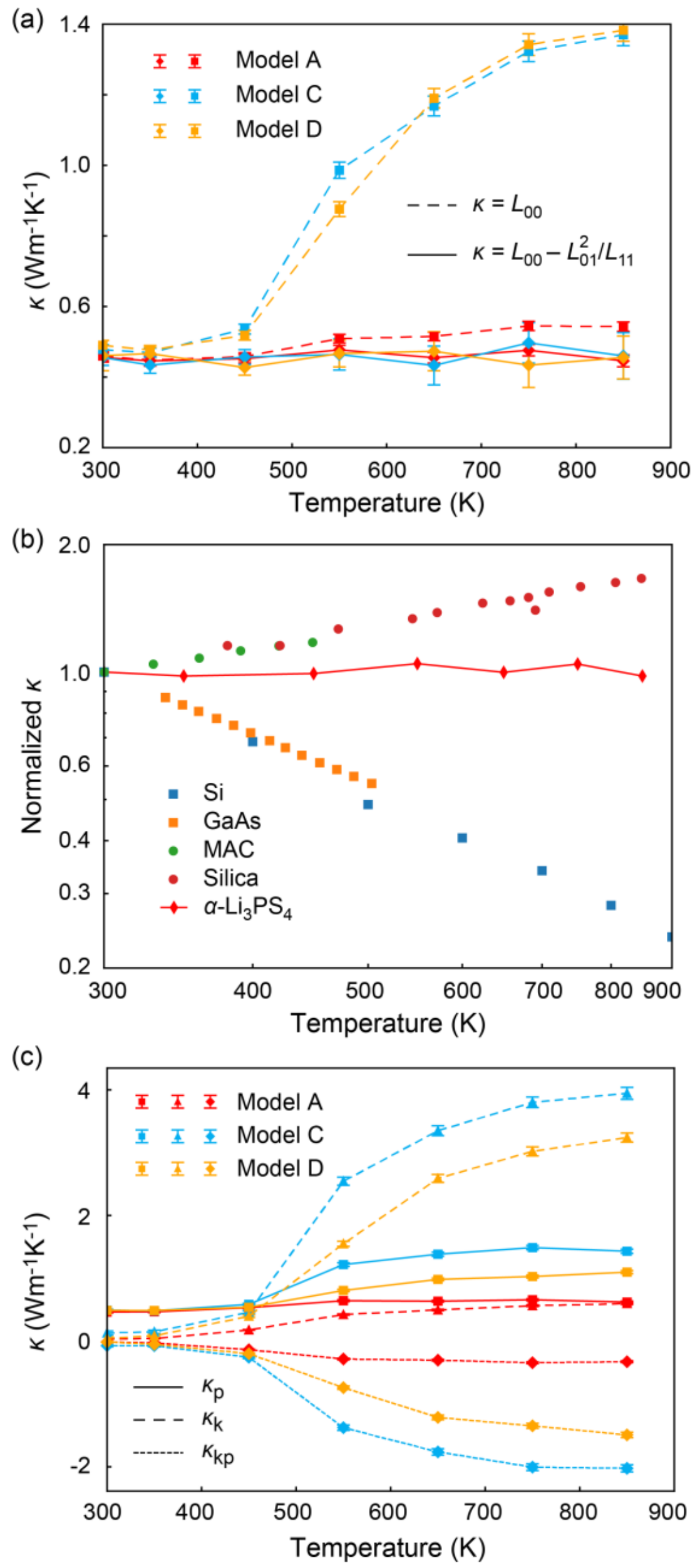

**FIG. 3. Thermal conductivity of $\alpha$-$Li_3PS_4$ and its decomposition based on energy flux as a function of temperature.** (a) Thermal conductivity calculated from the Green-Kubo integral based on $L_{00}$ (dashed lines) and also $L_{00} - L_{01}^2/L_{11}$ (solid lines) via the machine-learned models A (red), C (blue), and D (yellow). (b) Temperature-dependent $\kappa$ of $\alpha$-$Li_3PS_4$ in comparison with representative crystals and glasses including silicon (Si) [62], gallium arsenide (GaAs) [61], silica [64], and monolayer amorphous carbon (MAC) [63]. All the datasets have been normalized by their values at 300 K. (c) Energy flux-based decomposition of the thermal conductivity ($L_{00}$) into $\kappa_p$, $\kappa_k$, and $\kappa_{kp}$, which represent contributions of the potential (solid), kinetic (dashed), and cross (dotted) terms, respectively. Both $\kappa_p$ and $\kappa_k$ remain positive while $\kappa_{kp}$ is always negative [33].

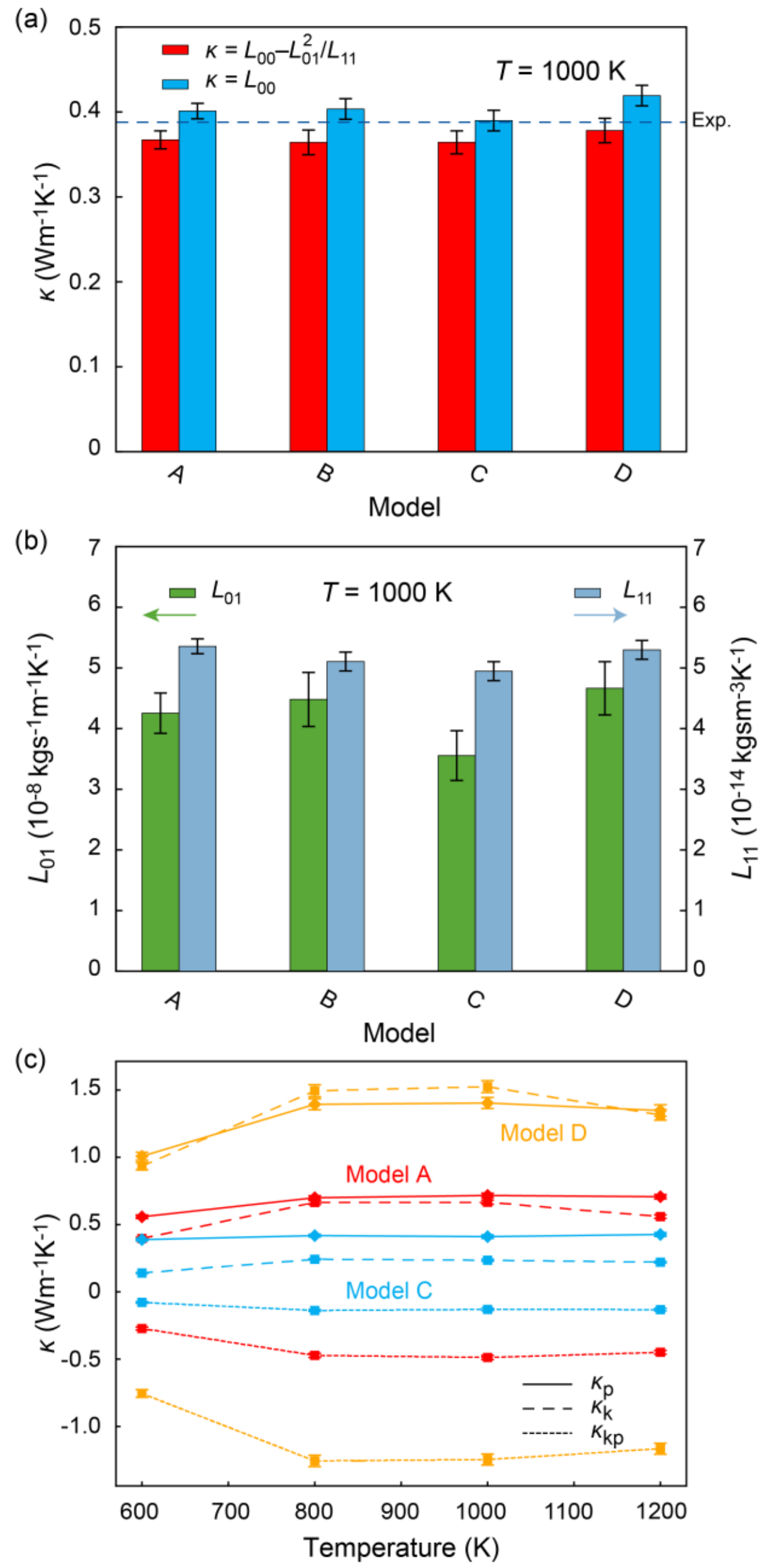

**FIG. 4. Thermal conductivity and Onsager coefficients of $\beta$-$Cu_{1.98}$Se.** (a) Thermal conductivity at 1000 K obtained from the direct Green-Kubo integral of energy flux $L_{00}$ (blue) and the formula $L_{00} - L_{01}^2/L_{11}$ (red) using different machine-learned models. Experimentally measured thermal conductivity at 1000 K is also plotted (blue dashed line) [14], where the electronic contribution has been subtracted for direct comparison. (b) $L_{01}$ and $L_{11}$ of $\beta$-$Cu_{1.98}$Se at 1000 K. (c) Thermal conductivity $L_{01}$ decomposition into $\kappa_p$, $\kappa_k$, and $\kappa_{kp}$ based on the energy flux as a function of temperature.

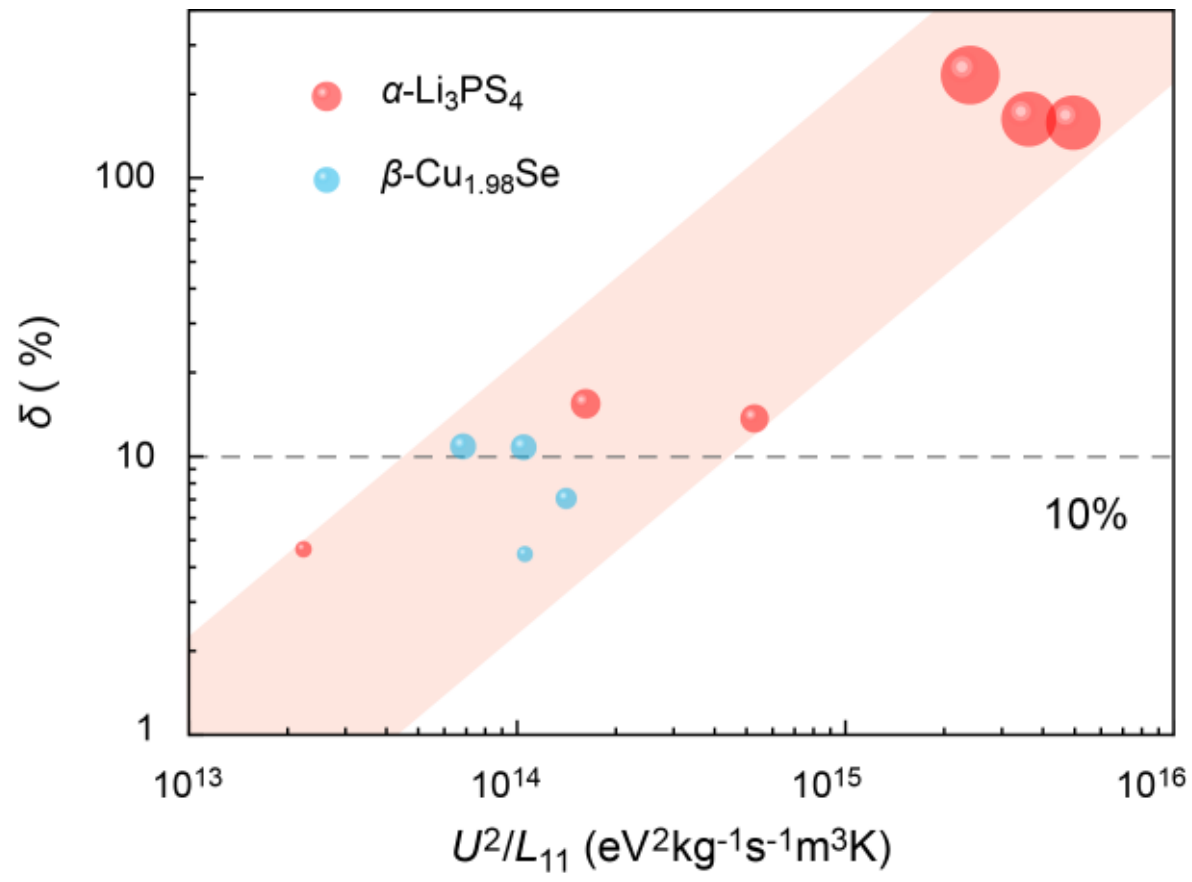

**FIG. 5. Relative error of thermal conductivity calculated using conventional GK approach.** Here, $\delta = (\kappa_{L_{00}} - \kappa_{L_{00}-L_{01}^2/L_{11}})/\kappa_{L_{00}-L_{01}^2/L_{11}} \times 100\%$. All data points are extracted from Fig. 2(a) and Fig. 4(a). The dashed line marks the 10% threshold. The symbol size reflects the magnitude of $\delta$ on a logarithmic scale.